\documentclass[11pt,superscriptaddress,aps,prd,preprint]{revtex4}
\usepackage{amsmath}

\makeatletter
\usepackage[T1]{fontenc}
\usepackage{amsmath}
\usepackage{amssymb}
\usepackage{graphicx}

\usepackage{slashed}

\newcommand{\bea}{\begin{eqnarray}}
\newcommand{\eea}{\end{eqnarray}}

\begin{document}

\title{Lorentz violation and Gravitoelectromagnetism: Casimir effect and Stefan-Boltzmann law at Finite temperature}

\author{A. F. Santos}\email[]{alesandroferreira@fisica.ufmt.br}
\affiliation{Instituto de F\'{\i}sica, Universidade Federal de Mato Grosso,\\
78060-900, Cuiab\'{a}, Mato Grosso, Brazil}

\author{Faqir C. Khanna\footnote{Professor Emeritus - Physics Department, Theoretical Physics Institute, University of Alberta\\
Edmonton, Alberta, Canada}}\email[]{khannaf@uvic.ca}
\affiliation{Department of Physics and Astronomy, University of Victoria,\\
3800 Finnerty Road Victoria, BC, Canada}

\begin{abstract}
 
The standard model and general relativity are local Lorentz invariants. However it is possible that at Planck scale there may be a breakdown of Lorentz symmetry. Models with Lorentz violation are constructed using Standard Model Extension (SME). Here gravitational sector of the SME is considered to analyze the Lorentz violation in Gravitoelectromagnetism (GEM). Using the energy-momentum tensor, the Stefan-Boltzmann law and Casimir effect are calculated at finite temperature  to ascertain the level of local Lorentz violation. Thermo Field Dynamics (TFD) formalism is used to introduce temperature effects.

\end{abstract}



\maketitle

\section{Introduction}

Lorentz and CPT symmetries play a central role in the Standard Model (SM) and Einstein General Relativity (GR). GR is a classical theory that describes the gravitational force. The SM describes other three fundamental forces that are defined in a quantum version. There are models that seek to unify the two fundamental theories into a single one. Such a theory is expected to emerge at the Planck scale, $\sim 10^{19}\mathrm{GeV}$, where some new physics may emerge. The new physics may involve different properties, such as the appearance of Lorentz violation effects \cite{Kost1989, Kost1991}. Studies of Lorentz violation, both theoretical and experimental, are described by an effective field theory called the Standard Model Extension (SME)  \cite{SME1}. The SME includes the SM, GR and all possible operators that break the Lorentz symmetry. A complete description of GR in the framework of the SME has been considered \cite{Kos0, Kos01, Kos02}. In the gravitational sector of the SME \cite{Hees, Iorio2} there are 19 coefficients for Lorentz violation in addition to an unobservable scalar parameter. A similarity between the gravitational sector and the electromagnetic sector of the SME, specifically CPT-even coefficients, has been developed \cite{QG}. This would suggest a close relationship between gravitational and CPT-even electromagnetic sectors.

The search for analogies between electromagnetism and gravity, for Lorentz invariant theories, started with Faraday \cite{Faraday} and Maxwell \cite{Maxwell} and has a long history \cite{Heaviside1, Weyl, KK, Thirring, Matte, Campbell1, Campbell2, Campbell4, Braginsky}. For a review of Gravitoelectromagnetism (GEM) follow references \cite{Mashhon}. Experimental efforts to test GEM have been developed \cite{Iorio4}. There are three different ways to construct GEM theory: (i) using the similarity between the linearized Einstein and Maxwell equations  \cite{Mashhon}; (ii) based on an approach using tidal tensors  \cite{Filipe} and (iii) the decomposition of the Weyl tensor into gravitomagnetic (${\cal B}_{ij}=\frac{1}{2}\epsilon_{ikl}C^{kl}\,_{0j}$) and gravitoelectric (${\cal E}_{ij}=-C_{0i0j}$) approach \cite{Maartens}. Here Weyl approach is considered. The Weyl tensor is connected with the curvature tensor and it is the trace-less part of the Riemann tensor. The analogy between electromagnetism and General Relativity is based on the correspondence  $C_{\alpha\sigma\mu\nu} \leftrightarrow  F_{\alpha\sigma}$, where the Weyl tensor is the free gravitational field and $F_{\alpha\sigma}$ is the electromagnetic tensor. The Weyl tensor gives contributions due to nonlocal sources. In the Weyl tensor approach, a Lagrangian formulation for GEM has been developed \cite{Khanna}. In this formalism a symmetric gravitoelectromagnetic tensor potential, $A_{\mu\nu}$, which describes the gravitational interaction, is defined. For example, the GEM theory at finite temperature has been analyzed \cite{our1}. The gravitational Bhabha scattering has been calculated \cite{our2}. Using the Lagrangian formalism for GEM, our main objective in this paper is to calculate contributions of the Lorentz violation to the Casimir effect and the Stefan-Boltzmann law of the GEM theory.

The Casimir effect is the interaction between two parallel conducting plates \cite{Casimir}. The attraction between plates is the result of electromagnetic modes due to boundary conditions or topological effects. Initially this effect was predicted for the electromagnetic field. However now it has been defined for any quantum field. The Casimir effect was confirmed experimentally first by Sparnaay \cite{Sparnaay}. Now high degree of accuracy has been achieved experimentally \cite{Lamoreaux}, \cite{Mohideen}. If the gravitational field has a quantum nature, this effect would be expected for gravitational waves. Using the GEM formulation and considering plates that are made of superconducting material, the gravitational Casimir effect has been analyzed \cite{Quach}. The Casimir effect for GEM at finite temperature has been calculated \cite{our}. In the present study the Casimir energy and pressure and the Stefan-Boltzmann law for the GEM field with Lorentz-violating corrections at finite temperature are calculated. The Thermo Field Dynamics (TFD) formalism is used to introduce the finite temperature effects.

TFD is a real-time finite temperature formalism \cite{Umezawa1, Umezawa2}. This formalism leads to an interpretation of the statistical average of an arbitrary operator ${\cal O}$, as the expectation value in a thermal vacuum, i.e., $\langle {\cal O} \rangle=\langle 0(\beta)| {\cal O}|0(\beta) \rangle$. The thermal vacuum $|0(\beta) \rangle$ describes the thermal equilibrium of the system, where  $\beta=\frac{1}{k_BT}$, $T$ is the temperature and $k_B$ is the Boltzmann constant. To construct this thermal state two basic elements are necessary: (i) the doubling of the original Fock space and (ii) the Bogoliubov transformation. This doubling consists of Fock space composed of the original, $S$, and a fictitious space (tilde space), $\tilde{S}$. The map between the tilde and non-tilde operators is defined by the tilde (or dual) conjugation rules. The Bogoliubov transformation is a rotation among operators involving these two spaces. Here we use natural units, i.e., $k_B=\hbar=c=1$.

This paper is organized as follows. In section II, a Lagrangian formulation for GEM is introduced. In section III, the GEM theory with Lorentz-violating parameter is analyzed. The vacuum expectation value of the energy-momentum tensor is calculated. In section IV, TFD and some characteristics of the finite temperature formalism are presented. In section V, some applications are developed. The Stefan-Boltzmann law and the Casimir effect with Lorentz-violating corrections at zero and finite temperature are calculated. In section VI, some concluding remarks are presented.

\section{An introduction to GEM field}

A brief introduction to the lagrangian formulation of GEM is presented in this section. The GEM describes the dynamics of the gravitational field in a manner similar to that of the electromagnetic field. Here the GEM approach will be used with the Weyl tensor components ($C_{ijkl}$) being: ${\cal E}_{ij}=-C_{0i0j}$ (gravitoelectric field) and ${\cal B}_{ij}=\frac{1}{2}\epsilon_{ikl}C^{kl}_{0j}$ (gravitomagnetic field). The field equations for the components of the Weyl tensor have a structure similar to those of Maxwell equations. The GEM equations are given as
\bea
\partial^i{\cal E}^{ij}&=&-4\pi G\rho^j,\label{01}\\ 
\partial^i{\cal B}^{ij}&=&0,\label{02}\\
\epsilon^{( i|kl}\partial^k{\cal B}^{l|j)}+\frac{\partial{\cal E}^{ij}}{\partial t}&=&-4\pi G J^{ij},\label{03}\\ 
\epsilon^{( i|kl}\partial^k{\cal E}^{l|j)}+\frac{\partial{\cal B}^{ij}}{\partial t}&=&0,\label{04}
\eea
where $G$ is the gravitational constant, $\epsilon^{ikl}$ is the Levi-Civita symbol, $\rho^j$ is the vector mass density and $J^{ij}$ is the mass current density. The symbol $(i|\cdots|j)$ denotes symmetrization of the first and last indices, i.e., $i$ and $j$. 

A lagrangian formulation for the GEM equations has been constructed \cite{Khanna}. In such a construction, the fields ${\cal E}^{ij}$ and ${\cal B}^{ij}$ are defined as 
\bea
{\cal E}=-\mathrm{grad}\,\varphi-\frac{\partial \tilde{\cal A}}{\partial t},\quad\quad\quad\quad {\cal B}=\mathrm{curl}\,\tilde{\cal A},
\eea
where $\tilde{\cal A}$ with components ${\cal A}^{\mu\nu}$ is a symmetric rank-2 tensor field, gravitoelectromagnetic tensor potential,  and $\varphi$ is the GEM vector counterpart of the electromagnetic scalar potential $\phi$. A gravitoelectromagnetic tensor $F^{\mu\nu\alpha}$ is defined as
\bea
F^{\mu\nu\alpha}=\partial^\mu{\cal A}^{\nu\alpha}-\partial^\nu{\cal A}^{\mu\alpha},
\eea
where $\mu, \nu,\alpha=0, 1, 2, 3$. Then GEM equations are written as
\bea
\partial_\mu{F}^{\mu\nu\alpha}=4\pi G{\cal J}^{\nu\alpha},\quad\quad\quad\quad\partial_\mu{\cal G}^{\mu\langle\nu\alpha\rangle}=0,
\eea
where ${\cal J}^{\nu\alpha}$ depends on the mass density ($\rho^i$) and the current density ($J^{ij}$) and ${\cal G}^{\mu\nu\alpha}$ is the dual GEM tensor defined as ${\cal G}^{\mu\nu\alpha}=\frac{1}{2}\epsilon^{\mu\nu\gamma\sigma}\eta^{\alpha\beta}{F}_{\gamma\sigma\beta}.$
Then the GEM lagrangian is 
\bea
{\cal L}_G=-\frac{1}{16\pi}{F}_{\mu\nu\alpha}{F}^{\mu\nu\alpha}-G\,{\cal J}^{\nu\alpha}{\cal A}_{\nu\alpha}.\label{L_G}
\eea
Since the nature of $A_{\mu\nu}$ is different from $h_{\mu\nu}$, we use a different approach. The tensor potential is not related to the perturbation of the spacetime.
It is connected directly with the description of the gravitational field in flat spacetime. The gauge transformation for the tensor potential is $A'_{\mu\nu}=A_{\mu\nu}+\partial_\mu\theta_\nu$, where $\theta_\nu$ is 4-vector. The gravitoelectromagnetic tensor $F_{\mu\nu\alpha}$ is invariant under this transformation. Then the GEM Lagrangian is gauge invariant. For more details see \cite{Ramos}. Therefore, the gauge transformation in GEM is similar to that of electromagnetism.

\section{Lorentz-violating contributions to the GEM field}

The Lagrangian that includes the  Lorentz-violating contributions to the GEM field is 
\bea
{\cal L}=-\frac{1}{16\pi}F_{\rho\sigma\theta}F^{\rho\sigma\theta}-\frac{1}{4}\bigl(k^{(4)}\bigl)_{\kappa\lambda\xi\rho}\,\eta_{\gamma\theta}F^{\kappa\lambda\gamma}F^{\xi\rho\theta},\label{Lag}
\eea
where $\bigl(k^{(4)}\bigl)_{\kappa\lambda\xi\rho}$ is a dimensionless coefficient field that belongs to minimal sector of SME gravity \cite{Xu, Kos0}. This tensor has the same symmetries as the Riemann tensor and can be decomposed into 20 coefficients, i.e., $s^{\mu\nu}$ with 9 independent quantities, $t^{\mu\nu\alpha\gamma}$ that have symmetries of the Riemann curvature tensor, implying 10 independent quantities and $u$ is a scalar. In the weak field approximation, coefficients for Lorentz violation are taken as constants in a special coordinate system and are donated by $\bar{s}^{\mu\nu}$, $\bar{t}^{\mu\nu\alpha\gamma}$  and $\bar{u}$. The $\bar{u}$ coefficient is not observable. Then the gravitational sector has 19 coefficients. The CPT-even part of the electromagnetic (EM) field sector has 19 Lorentz violation coefficients which are decomposed into two parts: 10 birefringent and 9 non-birefringent components. In addition, coefficients of the gravity sector are reminiscent of those for the coefficient $\left(k_F\right)_{\mu\nu\alpha\beta}$ in the electromagnetic part of the SME \cite{Kos0}. Therefore, there is a correspondence between Lorentz violation effects for the EM field and for the weak field gravitational field, i.e. GEM field \cite{QG}.  Here, for simplicity, the calculations are developed considering all components of the tensor $\bigl(k^{(4)}\bigl)_{\kappa\lambda\xi\rho}$.

In order to calculate the Casimir effect, first the energy-momentum tensor for the Lagrangian (\ref{Lag}) is defined as
\bea
\mathbb{T}^{\mu\nu}&=&\frac{\partial {\cal L}}{\partial(\partial_\mu A_{\lambda\xi})}\partial^\nu A_{\lambda\xi}-g^{\mu\nu}{\cal L}.
\eea
Here, this tensor is divided into two parts,
\bea
\mathbb{T}^{\mu\nu}&=&\mathbb{T}^{\mu\nu}_{GEM}+\mathbb{T}^{\mu\nu}_{LV}, \label{EMT1}
\eea
where
\bea
\mathbb{T}^{\mu\nu}_{GEM}=-\frac{1}{4\pi}{\cal F}^{\mu\lambda\xi}\partial^\nu A_{\lambda\xi}+\frac{1}{16\pi}g^{\mu\nu}{\cal F}_{\rho\sigma\theta}{\cal F}^{\rho\sigma\theta}
\eea
is the part that corresponds to the GEM field and 
\bea
\mathbb{T}^{\mu\nu}_{LV}&=&-\bigl(k^{(4)}\bigl)^{\kappa\lambda\mu\xi}F_{\kappa\lambda}\,^{\Lambda}\partial^\nu A_{\xi\Lambda}+\frac{1}{4}g^{\mu\nu}\bigl(k^{(4)}\bigl)_{\kappa\lambda\xi\rho} F^{\kappa\lambda}\,_\theta F^{\xi\rho\theta}
\eea
is the Lorentz-violating part. It is to be noted that this tensor (\ref{EMT1}) is not symmetric. The Belinfante method \cite{belif} is used to define the Lorentz invariant part. Then the symmetric energy-momentum is
\bea
T^{\mu\nu}_{GEM}=\frac{1}{4\pi}\left[-{\cal F}^\mu\,_{\lambda\xi}{\cal F}^{\nu\lambda\xi}+\frac{1}{4}g^{\mu\nu}{\cal F}_{\rho\sigma\theta}{\cal F}^{\rho\sigma\theta}\right].\label{EMT2}
\eea
The same method is not applicable for the Lorentz violating part. However this is written as
\bea
T^{\mu\nu}_{LV}&=&-\bigl(k^{(4)}\bigl)^{\kappa\lambda\mu\rho}F^\nu\,_{\rho\Lambda}F_{\kappa\lambda}\,^\Lambda+\frac{1}{4}g^{\mu\nu}\bigl(k^{(4)}\bigl)_{\kappa\lambda\xi\rho} F^{\kappa\lambda}\,_\theta F^{\xi\rho\theta}.
\eea
Thus the total energy-momentum tensor becomes
\bea
T^{\mu\nu}=\frac{1}{4\pi}\left[-{\cal F}^\mu\,_{\lambda\xi}{\cal F}^{\nu\lambda\xi}+\frac{1}{4}g^{\mu\nu}{\cal F}_{\rho\sigma\theta}{\cal F}^{\rho\sigma\theta}\right]-\bigl(k^{(4)}\bigl)^{\kappa\lambda\mu\rho}F^\nu\,_{\rho\gamma}F_{\kappa\lambda}\,^\gamma +\frac{1}{4}g^{\mu\nu}\bigl(k^{(4)}\bigl)_{\kappa\lambda\xi\rho} F^{\kappa\lambda}\,_\theta F^{\xi\rho\theta}.
\eea
This tensor is not completely symmetric. This is a feature of theories which exhibit Lorentz violation.

The canonical conjugate momentum related to the tensor $A_{\kappa\lambda}$ is given as
\bea
\pi^{\kappa\lambda}=\frac{\partial {\cal L}}{\partial(\partial_0 A_{\kappa\lambda})}=-\frac{1}{4\pi}{\cal F}^{0\kappa\lambda}.
\eea
Adopting the Coulomb gauge, where $A^{0i}=0$ and $\mathrm{div}\tilde{A}=\partial_i A^{ij}=0$, the covariant quantization is carried out and the commutation relation is
{\small
\bea
\left[A^{ij}({\bf x},t),\pi^{kl}({\bf x}',t)\right]=\frac{i}{2}\Bigl[\delta^{ik}\delta^{jl}-\delta^{il}\delta^{jk}-\frac{1}{\nabla^2}\Bigl(\delta^{jl}\partial^i\partial^k-\delta^{jk}\partial^i\partial^l-\delta^{il}\partial^j\partial^k+\delta^{ik}\partial^j\partial^l\Bigl)\Bigl]\delta^3({\bf x}-{\bf x}').\label{CR}
\eea}
Other commutation relations are zero.

To avoid divergences, the energy-momentum tensor is written at different space-time points as
\bea
T^{\mu\nu}(x)=T^{\mu\nu}_{GEM}(x)+T^{\mu\nu}_{LV}(x),
\eea
where
\bea
T^{\mu\nu}_{GEM}(x)&=&\frac{1}{4\pi}\lim_{x'\rightarrow x}\Bigl[-\mathbb{F}^\mu\,_{\lambda\xi,}\,^{\nu\lambda\xi}(x,x')+\frac{1}{4}g^{\mu\nu}\mathbb{F}^{\rho\sigma\theta,}\,_{\rho\sigma\theta}(x,x')\Bigl],
\eea
and
\bea
T^{\mu\nu}_{LV}(x)&=&\lim_{x'\rightarrow x}\Bigl[-\bigl(k^{(4)}\bigl)^{\kappa\lambda\mu\rho}\mathbb{F}^\nu\,_{\rho\gamma},_{\kappa\lambda}\,^\gamma(x,x')+\frac{1}{4}g^{\mu\nu}\bigl(k^{(4)}\bigl)_{\kappa\lambda\xi\rho} \mathbb{F}^{\kappa\lambda}\,_\theta,^{\xi\rho\theta}(x,x')\Bigl],
\eea
where
\bea
\mathbb{F}^{\xi\kappa\gamma,\mu\nu\rho}(x,x')\equiv\tau\left[{\cal F}^{\xi\kappa\gamma}(x){\cal F}^{\mu\nu\rho}(x')\right].
\eea
with $\tau$ being the time order operator. Using the $\tau$ operator explicity
\bea
\mathbb{F}^{\xi\kappa\gamma,\mu\nu\rho}(x,x')&=&{\cal F}^{\xi\kappa\gamma}(x){\cal F}^{\mu\nu\rho}(x')\theta(x_0-x_0')+{\cal F}^{\mu\nu\rho}(x'){\cal F}^{\xi\kappa\gamma}(x)\theta(x_0'-x_0),
\eea
with $\theta(x_0-x_0')$ being the step function. In calculations, the commutation relation eq. (\ref{CR}) and 
\bea
\partial^\mu\theta(x_0-x_0')=n^\mu_0\delta(x_0-x_0'),
\eea
 are used where $n^\mu_0=(1,0,0,0)$ is a time-like vector. Then we get
\bea
\mathbb{F}^{\xi\kappa\gamma,\mu\nu\rho}(x,x')&=&\Gamma^{\xi\kappa\gamma,\mu\nu\rho,\lambda\epsilon\omega\upsilon}(x,x')\tau\left[A_{\lambda\epsilon}(x)A_{\omega\upsilon}(x')\right]\nonumber\\
&+&I^{\kappa\gamma, \mu\nu\rho}(x,x')n^\xi_0 \delta(x_0-x_0')-I^{\xi\gamma, \mu\nu\rho}(x,x')n^\kappa_0 \delta(x_0-x_0'),
\eea
where 
\bea
\Gamma^{\xi\kappa\gamma,\mu\nu\rho,\lambda\epsilon\omega\upsilon}(x,x')&=&\left(g^{\kappa\lambda}g^{\epsilon\gamma}\partial^\xi-g^{\xi\lambda}g^{\epsilon\gamma}\partial^\kappa\right)\left(g^{\nu\omega}g^{\rho\upsilon}\partial'^\mu-g^{\mu\omega}g^{\rho\upsilon}\partial'^\nu\right)
\eea
and
\bea
I^{\kappa\gamma, \mu\nu\rho}(x,x')&=&\left[A^{\kappa\gamma}(x), {\cal F}^{\mu\nu\rho}(x')\right]
\eea

Then the complete energy-momentum tensor is
\bea
T^{\mu\nu}(x)&=&-\lim_{x'\rightarrow x}\Bigl\{\Bigl(\frac{1}{4\pi}\Delta^{\mu\nu,\lambda\epsilon\omega\upsilon}_{GEM}(x,x')+\Delta^{\mu\nu,\lambda\epsilon\omega\upsilon}_{LV}(x,x')\Bigl)\tau\left[A_{\lambda\epsilon}(x)A_{\omega\upsilon}(x')\right]\Bigl\},
\eea
with
\bea
\Delta^{\mu\nu,\lambda\epsilon\omega\upsilon}_{GEM}(x,x')&=&\Gamma^\mu\,_{\rho\xi,}\,^{\nu\rho\xi,\lambda\epsilon\omega\upsilon}(x,x')-\frac{1}{4}g^{\mu\nu}\Gamma^{\rho\sigma\theta,}\,_{\rho\sigma\theta,}\,^{\lambda\epsilon\omega\upsilon}(x,x')
\eea
and 
\bea
&&\Delta^{\mu\nu,\lambda\epsilon\omega\upsilon}_{LV}(x,x')=\bigl(k^{(4)}\bigl)^{\kappa\lambda\mu\rho}\Gamma^\nu\,_{\rho\gamma,\kappa\lambda}\,^{\gamma,\lambda\epsilon\omega\upsilon}(x,x')-\frac{1}{4}g^{\mu\nu}\bigl(k^{(4)}\bigl)^{\kappa\lambda\gamma\rho} \Gamma_{\kappa\lambda\theta,}\,_{\gamma\rho}\,^{\theta,\lambda\epsilon\omega\upsilon}(x,x').
\eea

The vacuum expectation value of $T^{\mu\nu}$ is
\bea
\langle T^{\mu\nu}(x)\rangle &=&-\lim_{x'\rightarrow x}\Bigl\{\Bigl(\frac{1}{4\pi}\Delta^{\mu\nu,\lambda\epsilon\omega\upsilon}_{GEM}(x,x')+\Delta^{\mu\nu,\lambda\epsilon\omega\upsilon}_{LV}(x,x')\Bigl)\langle 0|\tau\left[A_{\lambda\epsilon}(x)A_{\omega\upsilon}(x')\right]|0\rangle\Bigl\}.\nonumber
\eea
Using the graviton propagator,
\bea
D_{\lambda\epsilon\omega\upsilon}(x-x')&=&\frac{i}{2}N_{\lambda\omega\epsilon\upsilon}\,G_0(x-x'),
\eea
where $N_{\lambda\omega\epsilon\upsilon}\equiv\eta_{\lambda\omega}\eta_{\epsilon\upsilon}+\eta_{\lambda\upsilon}\eta_{\epsilon\omega}-\eta_{\lambda\epsilon}\eta_{\omega\upsilon}$ and $G_0(x-x')$ is the massless scalar field propagator. An important note, Lorentz-violating coefficients are small and hence can be treated perturbatively. Thus, to obtain first-order corrections in Lorentz-violating coefficients, an expansion of the propagator is considered. By taking the zeroth-order term in Lorentz-violating parameter the vacuum expectation value of $T^{\mu\nu}$ is 
\bea
\langle T^{\mu\nu}(x)\rangle &=& -\frac{i}{2}\lim_{x'\rightarrow x}\Bigl\{\Bigl(\frac{1}{4\pi}\Gamma^{\mu\nu}_{GEM}+\Gamma^{\mu\nu}_{LV}\Bigl)G_0(x-x')\Bigl\},
\eea
where
\bea
\Gamma^{\mu\nu}_{GEM}(x,x')=8\left(\partial^\mu\partial'^\nu-\frac{1}{4}g^{\mu\nu}\partial^\rho\partial'_\rho\right).\label{Gamma}
\eea
and
\bea
&&\Gamma^{\mu\nu}_{LV}(x,x')=8\Bigl[\bigl(k^{(4)}\bigl)^{\kappa\lambda\mu}\,_{\lambda}\partial^\nu\partial'_\kappa+\bigl(k^{(4)}\bigl)^{\nu\lambda\mu\rho}\partial_\rho\partial'_\lambda\nonumber\\
&-&\frac{1}{4}g^{\mu\nu}\left(\bigl(k^{(4)}\bigl)^{\kappa\lambda\gamma}\,_\lambda\partial_\kappa\partial'_\gamma+\bigl(k^{(4)}\bigl)^{\lambda\kappa}\,_{\lambda}\,^\rho\partial_\kappa\partial'_\rho\right)\Bigl].
\eea

Using the tilde conjugation rules, the vacuum average of $T^{\mu\nu}$ in terms of the $\alpha$-dependent fields is
\bea
\langle T^{\mu\nu(ab)}(x;\alpha)\rangle&=&-\frac{i}{2}\lim_{x'\rightarrow x}\Bigl\{\Bigl(\frac{1}{4\pi}\Gamma^{\mu\nu}_{GEM}+\Gamma^{\mu\nu}_{LV}\Bigl)G_0^{(ab)}(x-x';\alpha)\Bigl\},
\eea
with the $\alpha$-parameter being a compactification parameter defined by $\alpha=(\alpha_0,\alpha_1,\cdots\alpha_{D-1})$. Here a field theory on the topology $\Gamma_D^d=(\mathbb{S}^1)^d\times \mathbb{R}^{D-d}$ with $1\leq d \leq D$ is considered. Then any set of dimensions of the manifold $\mathbb{R}^{D}$ can be compactified, where the circumference of the $nth$ $\mathbb{S}^1$ is specified by $\alpha_n$. $D$ are the space-time dimensions and $d$ is the number of compactified dimensions.

The physical energy-momentum tensor is defined as
\bea
{\cal T}^{\mu\nu (ab)}(x;\alpha)=\langle T^{\mu\nu(ab)}(x;\alpha)\rangle-\langle T^{\mu\nu(ab)}(x)\rangle.
\eea
This definition describes a renormaliation procedure to obtain measurable physical quantities at finite temperature. Both the energy-momentum tensor at finite and zero temperature are divergent. Then by subtracting the energy-momentum tensor at zero temperature non-divergent results are obtained at finite temperature. With this procedure a measurable physical quantity is given by
\bea
{\cal T}^{\mu\nu (ab)}(x;\alpha)&=&-\frac{i}{2}\lim_{x'\rightarrow x}\Bigl\{\Bigl(\frac{1}{4\pi}\Gamma^{\mu\nu}_{GEM}+\Gamma^{\mu\nu}_{LV}\Bigl)\overline{G}_0^{(ab)}(x-x';\alpha)\Bigl\},\label{VEV}
\eea
with 
\bea
\overline{G}_0^{(ab)}(x-x';\alpha)=G_0^{(ab)}(x-x';\alpha)-G_0^{(ab)}(x-x').
\eea
The relevant component of the Fourier representation is $\overline{G}_0(x-x';\alpha)\equiv\overline{G}_0^{(11)}(x-x';\alpha)$ that is given by
\bea
\overline{G}_0(x-x';\alpha)&=&\int\frac{d^4k}{(2\pi)^4}e^{-ik(x-x')}v^2(k_\alpha;\alpha)\left[G_0(k)-G_0^*(k)\right].\label{GF}
\eea
where $v^2(k_\alpha;\alpha)$ is the generalized Bogoliubov transformation \cite{GBT} that is given as
\bea
v^2(k_\alpha;\alpha)&=&\sum_{s=1}^d\sum_{\lbrace\sigma_s\rbrace}2^{s-1}\sum_{l_{\sigma_1},...,l_{\sigma_s}=1}^\infty(-\eta)^{s+\sum_{r=1}^sl_{\sigma_r}}\exp\Bigl[{-\sum_{j=1}^s\alpha_{\sigma_j} l_{\sigma_j} k^{\sigma_j}}\Bigl],\label{BT}
\eea
where $d$ is the number of compactified dimensions, $\eta=1(-1)$ for fermions (bosons) and $\lbrace\sigma_s\rbrace$ denotes the set of all combinations with $s$ elements. In order to obtain physical conditions at finite temperature and spatial confinement, $\alpha_0$ has to be taken as a positive real number, while $\alpha_n$ for $n=1,2,\cdots, d-1$ must be pure imaginary of the form $iL_n$.

\section{Thermo Field Dynamics}

A brief introduction to Thermo Field Dynamics (TFD) is presented. TFD is a real-time finite temperature field theory. In this formalism the usual Fock space ${\cal S}$ of the system is doubled, such that the expanded space is ${\cal S}_T={\cal S}\otimes \tilde{\cal S}$, which is applicable to systems in  a thermal equilibrium state. This doubling is defined by the tilde ($^\thicksim$) conjugation rules, associating each operator in ${\cal S}$ to two operators in ${\cal S}_T$. 


Thermal effects are introduced through a Bogoliubov transformation that corresponds to a rotation in the tilde and non-tilde variables. For bosons this is given as
\bea
d(\alpha)&=&u(\alpha)d(k)-v(\alpha)\tilde d^\dagger(k),\\
\tilde d^\dagger(\alpha)&=&u(\alpha)\tilde d^\dagger(k)-v(\alpha) d(k),
\eea
where $(d^\dagger, \tilde{d}^\dagger)$ are creation operators, $(d, \tilde{d})$ are destruction operators, and the algebraic rules for thermal operators are
\bea
\left[d(k, \alpha), d^\dagger(p, \alpha)\right]&=&\delta^3(k-p),\nonumber\\
\left[\tilde{d}(k, \alpha), \tilde{d}^\dagger(p, \alpha)\right]&=&\delta^3(k-p),\label{ComB}
\eea
and other commutation relations are null. The quantities $u(\alpha)$ and $v(\alpha)$ are related to the Bose distribution function as $v^2(\alpha)=(e^{\alpha\omega}-1)^{-1}$ and  $u^2(\alpha)=1+v^2(\alpha)$. Here $\omega=\omega(k)$ and $\alpha=\beta$.

A doublet notation is defined by
\bea
\left( \begin{array}{cc} d(\alpha)  \\ \tilde d^\dagger(\alpha) \end{array} \right)={\cal B}(\alpha)\left( \begin{array}{cc} d(k)  \\ \tilde d^\dagger(k) \end{array} \right),
\eea
where ${\cal B}(\alpha)$ is the Bogoliubov transformation given as
\bea
{\cal B}(\alpha)=\left( \begin{array}{cc} u(\alpha) & -v(\alpha) \\ 
-v(\alpha) & u(\alpha) \end{array} \right).
\eea

As an example, let us consider a free scalar field in Minkowski space-time specified by $diag(g^{\mu\nu})=(+1,-1,-1,-1)$. The scalar field propagator is given as
\bea
G_0^{(ab)}(x-x';\alpha)=i\langle 0,\tilde{0}| \tau[\phi^a(x;\alpha)\phi^b(x';\alpha)]| 0,\tilde{0}\rangle,
\eea
where $\phi(x;\alpha)={\cal B}(\alpha)\phi(x){\cal B}^{-1}(\alpha)$ and $a, b=1,2$. Then
\bea
G_0^{(ab)}(x-x';\alpha)=i\int \frac{d^4k}{(2\pi)^4}e^{-ik(x-x')}G_0^{(ab)}(k;\alpha),
\eea
where 
\bea
G_0(k;\alpha)=G_0(k)+v^2(k;\alpha)[G_0(k)-G^*_0(k)],
\eea
with $G_0(k)=(k^2-m^2+i\epsilon)^{-1}$ and $[G_0(k)-G^*_0(k)]=2\pi i\delta(k^2-m^2)$. As the physical information is given by the non-tilde components, i.e. $G_0^{(11)}(k;\alpha)$, here $G_0^{(11)}(k;\alpha)\equiv G_0(k;\alpha)$ is used.

\section{Some applications}

Here three different applications which depend on the choice of the $\alpha$ parameter are considered.

\subsection{Stefan-Boltzmann law}

Consider the thermal effect for the choice $\alpha=(\beta,0,0,0)$. Then the generalized Bogoliubov transformation (\ref{BT}) becomes
\bea
v^2(\beta)=\sum_{j_0=1}^\infty e^{-\beta k^0 j_0}.
\eea
Then the Green function, eq. (\ref{GF}), is given as
\bea
\overline{G}_0^{(11)}(x-x';\alpha)=2\sum_{j_0=1}^\infty G_0\left(x-x'-i\beta j_0 n_0\right),\label{1GF}
\eea
where $n_0^\mu=(1,0,0,0)$. The vacuum expectation value of the energy-momentum tensor, eq. (\ref{VEV}), becomes
\bea
{\cal T}^{\mu\nu (11)}(x;\alpha)&=&-i\lim_{x'\rightarrow x}\Bigl\{\Bigl(\frac{1}{4\pi}\Gamma^{\mu\nu}_{GEM}+\Gamma^{\mu\nu}_{LV}\Bigl)\sum_{j_0=1}^\infty G_0\left(x-x'-i\beta j_0 n_0\right)\Bigl\}.
\eea
For $\mu=\nu=0$, we obtain 
\bea
{\cal T}^{00 (11)}(T)=\frac{\pi}{30}\left(1+4\pi \kappa_0\right)T^4,
\eea
the Stefan-Boltzmann law for the GEM field with corrections due to Lorentz-violating parameters, with
\bea
\kappa_0&\equiv& \frac{1}{2}\bigl(k^{(4)}\bigl)^{0\lambda 0}\,_\lambda + \bigl(k^{(4)}\bigl)^{0000}-\frac{1}{6}\Bigl(\bigl(k^{(4)}\bigl)^{1\lambda 1}\,_\lambda\nonumber\\
&+&\bigl(k^{(4)}\bigl)^{2\lambda 2}\,_\lambda+\bigl(k^{(4)}\bigl)^{3\lambda 3}\,_\lambda-2\bigl(k^{(4)}\bigl)^{0101}-2\bigl(k^{(4)}\bigl)^{0202}-2\bigl(k^{(4)}\bigl)^{0303}\Bigl).
\eea
When $\kappa_0=0$, the Lorentz invariant result obtained in \cite{our} is recovered. More results in statistical mechanics in the presence of Lorentz-violating background fields have been studied \cite{Colladay2004-06}.

The component $\mu=\nu=3$ is given as
\bea
{\cal T}^{33 (11)}(\beta)&=&\frac{\pi}{90\beta^4}\left(1+4\pi\kappa_1\right),
\eea
where
\bea
\kappa_1&\equiv& 3\bigl(k^{(4)}\bigl)^{0303}+\frac{3}{2}\bigl(k^{(4)}\bigl)^{0\lambda 0}\,_\lambda-\bigl(k^{(4)}\bigl)^{3\lambda 3}\,_\lambda +\bigl(k^{(4)}\bigl)^{3333}\nonumber\\
&+&\frac{5}{4}\left(\bigl(k^{(4)}\bigl)^{1\lambda 1}\,_\lambda+\bigl(k^{(4)}\bigl)^{2\lambda 2}\,_\lambda+\bigl(k^{(4)}\bigl)^{3\lambda 3}\,_\lambda\right).
\eea

It is important to observe that  the lowest order of the Lorentz violation leads to a modification in the Stefan-Boltzmann law. However, while small, Lorentz violating terms do not contradict any experimental measurements of the Stefan-Boltzmann law. In addition, constraints on the Lorentz-violating parameters can be obtained if the precision of the measurements will improve significantly.

\subsection{Casimir effect at zero temperature}

The Casimir effect for the GEM field of the SME with Lorentz symmetry violation at zero temperature is calculated. For parallel plates perpendicular to the z direction and separated by a distance $d$ the $\alpha$ parameter is chosen as $\alpha=(0,0,0,i2d)$. In this case, the Bogoliubov transformation is
\bea
v^2(d)=\sum_{l_3=1}^\infty e^{-i2dk^3l_3}
\eea
and the Green function is
\bea
\overline{G}_0^{(11)}(x-x';d)=2\sum_{l_3=1}^\infty G_0\left(x-x'-2dl_3z\right).\label{2GF}
\eea
Then the energy-momentum tensor becomes
\bea
{\cal T}^{\mu\nu (11)}(x;d)&=&-i\lim_{x'\rightarrow x}\Bigl\{\left(\frac{1}{4\pi}\Gamma^{\mu\nu}_{GEM}+\Gamma^{\mu\nu}_{LV}\right)\sum_{l_3=1}^\infty G_0\left(x-x'-2dl_3z\right)\Bigl\}.
\eea
Thus the Casimir energy and pressure are obtained
\bea
E(d)={\cal T}^{00 (11)}(d)&=&-\frac{\pi}{1440d^4}\left(1+4\pi \kappa_2\right)\\
P(d)={\cal T}^{33 (11)}(d)&=&-\frac{\pi}{480d^4}\left(1+4\pi \kappa_3\right),
\eea
where
\bea
\kappa_2&\equiv& \frac{1}{2}\bigl(k^{(4)}\bigl)^{0\lambda 0}\,_\lambda +\bigl(k^{(4)}\bigl)^{0000}+\frac{1}{2}\bigl(k^{(4)}\bigl)^{1\lambda 1}\,_\lambda\nonumber\\
&-&\bigl(k^{(4)}\bigl)^{0101}+\frac{1}{2}\bigl(k^{(4)}\bigl)^{2\lambda 2}\,_\lambda-\bigl(k^{(4)}\bigl)^{0202}-\frac{3}{2}\bigl(k^{(4)}\bigl)^{3\lambda 3}\,_\lambda + 3\bigl(k^{(4)}\bigl)^{0303}
\eea
and 
\bea
\kappa_3&\equiv& 3\bigl(k^{(4)}\bigl)^{0303}+\frac{3}{2}\bigl(k^{(4)}\bigl)^{0\lambda 0}\,_\lambda-\bigl(k^{(4)}\bigl)^{3\lambda 3}\,_\lambda\nonumber\\
&+&\bigl(k^{(4)}\bigl)^{3333}-\frac{5}{12}\Bigl(\bigl(k^{(4)}\bigl))^{1\lambda 1}\,_\lambda+\bigl(k^{(4)}\bigl)^{2\lambda 2}\,_\lambda-3\bigl(k^{(4)}\bigl)^{3\lambda 3}\,_\lambda\Bigl).
\eea
These expressions are consequences of the periodic conditions introduced by the topology $\Gamma_4^1=\mathbb{S}^1\times \mathbb{R}^{3}$ where $\mathbb{S}^1$ stands for the compactification of $x^3$-axis in a circumference of length $L=2d$. By taking $L=2d$ in the Green function is equivalent to the contributions of even images used in \cite{BM}, for Dirichlet boundary condition. Then the the toroidal topology method can be used for calculating the Casimir effect for Dirichlet boundary condition.

This result shows that the Lorentz-violating term modifies the Casimir effect for the GEM field. Since the Lorentz-violating parameter is small $(\kappa\ll 1)$, our result shows that the Casimir force between the plates is attractive, similar to the case of the electromagnetic field.

\subsection{Casimir effect at finite temperature}

The effect of finite temperature is introduced by taking $\alpha=(\beta, 0, 0,i2d)$ and then the Bogoliubov transformation, eq. (\ref{BT}), becomes
\bea
v^2(k^0,k^3;\beta,d)&=&\sum_{j_0=1}^\infty e^{-\beta k^0j_0}+\sum_{l_3=1}^\infty e^{-iLk^3l_3}+2\sum_{j_0,l_3=1}^\infty e^{-\beta k^0j_0-iLk^3l_3}.
\eea
The Green function, corresponding to the first two terms, is given in eq. (\ref{1GF}) and in eq. (\ref{2GF}), respectively. For the third term the Green function is
{\small
\bea
\overline{G}_0^{(11)}(x-x';\beta,d)=4\sum_{j_0,l_3=1}^\infty G_0\left(x-x'-i\beta j_0n-2dl_3z\right).\label{3GF}
\eea}
Then the energy-momentum tensor becomes
{\small
\bea
{\cal T}^{\mu\nu (11)}(\beta,d)&=&-2i\lim_{x'\rightarrow x}\Bigl\{\left(\frac{1}{4\pi}\Gamma^{\mu\nu}_{GEM}+\Gamma^{\mu\nu}_{LV}\right)\sum_{j_0,l_3=1}^\infty G_0\left(x-x'-i\beta j_0n-2dl_3z\right)\Bigl\}.
\eea}
The complete expression for the Casimir energy at finite temperature is obtained as
\bea
E(\beta,d)&=&\frac{\pi}{30\beta^4}\left(1+4\pi \kappa_0\right)-\frac{\pi}{1440d^4}\left(1+4\pi \kappa_2\right)\nonumber\\
&+&\frac{2}{\pi^3}\sum_{j_0,l_3=1}^\infty\frac{3(\beta j_0)^2-(2dl_3)^2}{[(\beta j_0)^2+(2dl_3)^2]^3}\left(1+4\pi \kappa_4\right),
\eea
where $E(\beta,d)={\cal T}^{00 (11)}(\beta,d)$. Here $\kappa_0, \kappa_2, \kappa_4$ represent the contribution of Lorentz violation into the Casimir Energy. The Casimir pressure at finite temperature is
\bea
P(\beta,d)&=&\frac{\pi}{90\beta^4}\left(1+4\pi\kappa_1\right)-\frac{\pi}{480d^4}\left(1+4\pi \kappa_3\right)\nonumber\\
&+&\frac{2}{\pi^3}\sum_{j_0,l_3=1}^\infty\frac{(\beta j_0)^2-3(2dl_3)^2}{[(\beta j_0)^2+(2dl_3)^2]^3}\left(1+4\pi \kappa_5\right),
\eea
where $P(\beta,d)={\cal T}^{33 (11)}(\beta,d)$. In this case $\kappa_1, \kappa_3, \kappa_5$ give the Lorentz violation effects to the Casimir
pressure. The first term is the Stefan-Boltzmann law, the second and third term are Casimir effect at zero and finite temperature, respectively. The Lorentz-violating parameters $\kappa_4$ and $\kappa_5$ are defined as
\bea
\kappa_4&\equiv&\frac{1}{2}\bigl(k^{(4)}\bigl)^{0\lambda 0}\,_\lambda+\bigl(k^{(4)}\bigl)^{0000}+\frac{3(2dl_3)^2-(\beta j_0)^2}{[(2dl_3)^2-3(\beta j_0)^2]}\left(-\frac{1}{2}\bigl(k^{(4)}\bigl)^{3\lambda 3}\,_\lambda+\bigl(k^{(4)}\bigl)^{0303}\right)\nonumber\\
&-&\frac{(\beta j_0)^2+(2dl_3)^2}{[(2dl_3)^2-3(\beta j_0)^2]}\Bigl(-\frac{1}{2}\bigl(k^{(4)}\bigl)^{1\lambda 1}\,_\lambda+\bigl(k^{(4)}\bigl)^{0101}-\frac{1}{2}\bigl(k^{(4)}\bigl)^{2\lambda 2}\,_\lambda+\bigl(k^{(4)}\bigl)^{0202}\Bigl)
\eea
and
\bea
\kappa_5&\equiv&\frac{(2dl_3)^2-3(\beta j_0)^2}{[3(2dl_3)^2-(\beta j_0)^2]}\left(\bigl(k^{(4)}\bigl)^{0303}+\frac{1}{2}\bigl(k^{(4)}\bigl)^{0\lambda 0}\,_\lambda\right)\nonumber\\
&+&\frac{1}{4}\bigl(k^{(4)}\bigl)^{3\lambda 3}\,_\lambda+\bigl(k^{(4)}\bigl)^{3333}-\frac{5}{4}\frac{(2dl_3)^2+(\beta j_0)^2}{[3(2dl_3)^2-(\beta j_0)^2]}\left(\bigl(k^{(4)}\bigl)^{1\lambda 1}\,_\lambda+\bigl(k^{(4)}\bigl)^{2\lambda 2}\,_\lambda\right).\nonumber
\eea
The modifications due to the Lorentz-violating terms at zero and finite temperature are similar.

These results are similar to the case of electromagnetic field. It is important to point out that although these results are similar there are important difference between two theories. For example, electromagnetic fields are vectors whereas GEM fields are tensors. The electromagnetic field propagates on a given space-time, whereas the gravitational field itself generates the space-time.

\section{Conclusions}

The SME is an effective theory that includes all Lorentz-violating parameter besides the known physics of the SM and GR. In this paper Lorentz-violating corrections to the GEM theory are considered. GEM is a gravitational theory based on an analogy with electromagnetism. A Lagrangian formalism of Gravitoelectromagnetism (GEM) is used. Using this formalism, the energy momentum tensor for the GEM field with Lorentz violation is calculated. Our main objective is to calculate the Lorentz- violating contributions to the Stefan-Boltzmann law and Casimir effect at finite temperature. The TFD formalism is used to introduce finite temperature effects. Our results show that contributions due to the Lorentz-violating term are linearly proportional to all components of the tensor $\bigl(k^{(4)}\bigl)_{\kappa\lambda\xi\rho}$.  Here the gravitational Casimir force is found to be proportional to $\sim(1+\kappa)F_G$, where $F_G$ is the gravitational Casimir effect and $\kappa$ is the correction due to the Lorentz violation. The gravitational Casimir effect for conventional plates is very small. However plates of special material, to measure the gravitational Casimir effect, using the GEM field have been developed \cite{Quach}. Thus, while small, Lorentz violating terms do not contradict any experimental measurements of the gravitational Casimir force and the Stefan-Boltzmann law. Our results indicate that the combined effect of temperature and compact space may, in principle, give a new constraint on the gravitational Casimir effect and Stefan-Boltzmann law as well as on the Lorentz violating parameters.

\acknowledgments
This work by A. F. S. is supported by CNPq projects 308611/2017-9 and 430194/2018-8.

\end{document}